\documentclass[epj,final]{svjour}
\usepackage[T1]{fontenc}
\usepackage[latin1]{inputenc}
\usepackage{graphics}
\usepackage{graphicx}
\makeatletter

\newcommand{\noun}[1]{\textsc{#1}}

\begin{document}

\title{Offsprings of a Point Vortex}

\author{Xavier Leoncini\inst{1} \and Alain Barrat\inst{1} \and Christophe Josserand\inst{2} \and  Simon Villain-Guillot\inst{3}}

\institute{Centre de Physique Théorique, Aix-Marseille Université, CNRS (UMR 6207), Luminy,
Case 907, F-13288 Marseille cedex 9, France \and Institut Jean le Rond d'Alembert, CNRS \& Université Pierre et Marie Curie (UMR 7190), Tour 55-65, 4 place Jussieu, 75252 Paris Cedex 05, France \and C.P.M.O.H. (UMR 5798), Université Bordeaux 1, 351 Cours de la Libération
33405 Talence Cedex, France}

\abstract{
The distribution engendered by successive splitting of one point vortex
are considered. The process of splitting a vortex in three using a
reverse three-point vortex collapse course is analysed in great details
and shown to be dissipative. A simple process of successive splitting
is then defined and the resulting vorticity distribution and vortex
populations are analysed.}
\maketitle

\section{Introduction}
Turbulent flows are often characterized by cascade dynamics that explain the transfer of quantities (energy, enstrophy)
between scales~\cite{frisch}. In particular, in two-dimensional turbulent flows, a striking features is
the presence of an inverse energy cascade, which leads to the emergence
of coherent vortices, dominating the flow dynamics \cite{Benzi86,Benzi88,McWilliams84,Weiss92,Carnevale91}, together with the direct cascade of enstrophy. 
In order to tackle these problems, point vortices have been commonly used with
some success:  indeed they can approximate the inviscid dynamics of finite-sized vortices
\cite{Zabusky82,Vobseek97,Fuentes96}, as for instance in punctuated
Hamiltonian models \cite{Carnevale91,Benzi92}. In these models, the
advection of well-separated vortices is approximated via Hamiltonian
point-vortex dynamics (thus inviscid {\it i.e.} dissipationless); to account for the change in the vortex population
toward smaller number of bigger vortices, dissipative merging processes
are then included for vortices which have approached each other closer
than a certain critical distance. 
In decaying two-dimensional turbulence, the merging process 
results in fact from the interaction of a few number of close vortices \cite{Dritschel96}
so that the understanding of low dimensional vortex dynamics is an essential
ingredient of the whole picture \cite{Aref99}. 
More generally, it has been shown that
point vortices can exhibit both the features of extremely high-dimensional
as well as low-dimensional systems \cite{Weiss98}. 

Since the pioneering work of Onsager on two dimensional turbulence
\cite{Onsager49}, a statistical approach to turbulence using point
vortices has also been developed. These vortex systems display negative
temperature, corresponding to states where same-sign vortices bound
to form larger vortices \cite{Joyce73,Onsager49}, although special care
 in the definition of the thermodynamic limit has to be done\cite{frohlich82}. In the
same spirit, stationary flows resulting from point vortices can be obtained~\cite{Caglioti92,Robert92,Spineanu05}
and different kinetic theories can also be derived (see for instance \cite{Chavanis07}
and references therein). 

Regarding the dynamical aspect, the relevance of point vortices, seen as exact solutions
of the Euler equation, is still debated: indeed finite time singularities are present in the coupled dynamics of many
point vortices. These singularities correspond to the collapse of three vortices \cite{Synge49,Novikov79,LKZ2000},
and can been seen as the consequences of an ill posed problem. Such singularity arises in fact naturally in the
Hamiltonian point vortices dynamics. Regardless the influence of the viscosity that would become dominant at short
time before the singularity, it is important to notice that the dynamics toward the three vortices collapses is in fact 
reversible. Thus, it is tempting to consider the reverse dynamics which would consist of vortices separation that has to be present in
this Hamiltonian system! Such dynamics has been omitted until now since the number of point vortices was taken constant or only decreasing.

Therefore, in this paper we take a different perspective on the existence of this
finite time singularity, using it as a potential source
of vorticity and vortices. One of the perspective is to offer
the possibility of a statistical mechanics approach with a varying
number of vortices. At this stage, we remain within the Hamiltonian framework of point vortices and in particular we do not consider the regularization due to the viscosity for real fluid. In addition, we focus here on the statistical distribution of vorticity rather than on the coupled dynamics of the system of vortices issued of this splitting. To be more precise, in this paper our main goal
is to describe the statistical properties emerging from this genuine process of vortex splitting: we consider the distributions of vortices and vortex
strength resulting from this simple mechanism of successive splitting according to the reverse
collapse of one point vortex. We shall refer to this result
as the offsprings of a point vortex. 

The paper is organized as follows,
first we recall briefly the notion of point vortices, and how they
naturally appear as a solution of Euler equation.  We then consider
the process of splitting one vortex in three. We introduce the relevant
parameters and analyse briefly the preliminary consequences. In particular, we observe that this splitting is inherently a dissipative process. Finally we define simple rules
for successive splitting. Computing
the offsprings of a point vortex with these rules we deduce general properties for the vortices distribution in the limit of high splitting processes.

\section{Basic equations}
 Point vortices are singular
solutions of some bidimensional physical systems described by a conservation
equation of what we shall call a generalised vorticity $\Omega$ given
by \begin{equation}
\frac{\partial\Omega}{\partial t}+\{\Omega,\psi\}=0\:,\label{vorti}\end{equation}
where $\{\cdot,\cdot\}$ denotes the usual Poisson bracket, and $\psi$
is a stream function. The actual relation $\Omega=F(\psi)$ may depend
on the considered physical system. For instance for the Euler equation
it is simply given by $\Omega=-\nabla^{2}\psi$. When $\Omega=-\nabla^{2}\psi+\psi/\rho_{s}^{2}$
where $\psi$ is, in this context, related to the electric potential
(in suitable units) in a plasma, $\rho_{s}$ is the hybrid Larmor
radius. Point vortices are defined by a vorticity distribution given
by a superposition of Dirac functions, \begin{equation}
\Omega(\mathbf{r},t)=\frac{1}{2\pi}\sum_{i=1}^{N}k_{i}\delta\left(\mathbf{r}-\mathbf{r}_{i}(t)\right)\:,\label{eq:Singular_vorticity}\end{equation}
where $\mathbf{r}=(x,y)$ is a vector in the plane of the flow, $k_{i}$
is the strength of vortex $i$ (circulation), $N$ is the total number
of vortices, and $\mathbf{r}_{i}(t)$ is the vortex position at time
$t$. Using this expression of the vorticity and solving the Poisson
equation, in the Euler case, or the Helmholtz equation in the more
general case, one obtains the current function associated to the point
vortices. Thanks to the Helmholtz theorem, the motion of the vortices is determined
by the value of the velocity field created by the other vortices at
the position of the vortex. The point vortex motion is Hamiltonian
and given by \begin{equation}
k_{i}\dot{y_{i}}=-\frac{\partial H}{\partial x_{i}}:,\hspace{10mm}\dot{x}_{i}=\frac{\partial H}{\partial(k_{i}y_{i})}:,(i=1,\cdots,N)\label{vortex.eq}\end{equation}
 where the Hamiltonian $H$ is given by \begin{equation}
H=\frac{1}{2\pi}\sum_{i>j}k_{i}k_{j}U(|\mathbf{r}_{i}-\mathbf{r}_{j}|)\label{ham}\end{equation}
with for an unbounded plane $U(x)=-\log(x)$ in the Euler case, and
$U(x)=\mathrm{K}_{0}(x)$ in the more general case of the plasma model (when $\rho_{s}\rightarrow\infty$
the modified Bessel function $\mathrm{K}_{0}$ tends to the logarithm). The Hamiltonian
(\ref{ham}) exhibits clearly that a system of point vortices is invariant
by translation and rotation, which implies both the conservation of
the centre of vorticity and the total angular momentum, the motion
of three vortices is integrable, while for four or more vortices Hamiltonian
chaos come into play. When the distance between the vortices is smaller
than the typical interaction length ($\rho_{s}$) the behaviour of
the two systems is similar, while in the large distance limit, the $\mathrm{K}_{0}$
interaction decreases exponentially and the vortices are almost free.
In the following, analytical computations will be made using the logarithmic
interaction, which corresponds to the Euler flow, and we can expect
the results to be qualitatively valid for the Bessel interaction as
long as the vortices are not {}``too far'' from each other.

Regarding the Hamiltonian (\ref{ham}), we remind  that it does not represent the 
energy of the fluid (which is infinite already with one point vortex), but rather corresponds to
an  energy of interaction between the vortices and is the one that is traditionnally used when 
making  a statistical physics approach of point vortex systems. It is however important to recall
 there is actually an infinite ``reserve'' of  energy in these systems. In what follows,
we abusively refer to this interaction-energy, as energy.

Let us now focus on a situation with only three vortices. In this
restricted situation, it is easier to tackle the motion of the vortices
by studying in fact their relative motion. Namely the
three vortices form a triangle, and the relative motion describes
the deformation of this triangle \cite{Synge49,Novikov75,Aref79,Aref83,LKZ2000}.
The invariance by translation of (\ref{ham}), allows us a free choice
of the origin of the plane, which we choose to be the centre of vorticity
(when it exists). The other constants of motion written in a frame
independent form become,

\begin{equation}
\left\{ \begin{array}{l}
H=-\frac{1}{2\pi}\left[k_{1}k_{2}\ln R_{3}+k_{1}k_{3}\ln R_{2}+k_{3}k_{2}\ln R_{1}\right]\\
K=k_{1}k_{2}R_{3}^{2}+k_{1}k_{3}R_{2}^{2}+k_{3}k_{2}R_{1}^{2}\:,\end{array}\right.\label{constantmotion1}\end{equation}
where $R_i=|\mathbf{r}_{j}-\mathbf{r}_{k}|$, with $i \ne j\ne k$.
In fact it has been known for a long time that the motion of vortices
can lead to singular solutions and finite time singularities, the
most striking one occurring for a system of three vortices resulting
in the collapse of the vortices in a finite time \cite{Synge49,Novikov79,LKZ2000}
or by time reversal, to an infinite expansion of the triangle formed
by the vortices. The collapse or infinite expansion of the three point
vortices are obtained when the following conditions are satisfied\begin{equation}
K=0\label{geomconditions}\end{equation}
 \begin{equation}
\sum_{i}\frac{1}{k_{i}}=0\:,\label{strengconditions}\end{equation}
 the harmonic mean of the vortex strengths (\ref{strengconditions})
and the total angular momentum in its frame free form (\ref{geomconditions}),
are both equal to zero \cite{Aref79,Aref83,Tavantzis88}.

In this paper we shall use this specific singularity and consider
vorticity distribution arising from successive splitting of a point
vortex according to the collapse conditions. Note that other type
of singularities involving more vortices are effectively possible,
however for the sake of simplicity, we restrict
ourselves to reverse three-vortex collapse rules. In what follows
we shall refer to the successive splitting process as the point vortex
offsprings.

\section{Splitting of a Point Vortex}

\subsection{Splitting rules}

In order to be consistent with the physical properties of vortex collapse,
we successively divide vortices according to the collapses rules and
keep the total vorticity constant. The splitting rules from one generation
$n$ of a vortex $i$ of strength $k_{i,n}$ to the next generation
$n+1$ read \begin{eqnarray}
\sum_{i=1}^{3}k_{i,n+1} & = & k_{i,n}\label{eq:Vorticity conservation}\\
\sum_{i=1}^{3}\frac{1}{k_{i,n+1}} & = & 0\:.\label{eq:Scale_free rule}\end{eqnarray}
These equations are equivalent to\begin{eqnarray}
\sum_{i=1}^{3}k_{i,n+1} & = & k_{i,n}\label{eq:plane}\\
\sum_{i=1}^{3}k_{i,n+1}^{2} & = & k_{i,n}^{2}\;,\label{eq:sphere}\end{eqnarray}
the $k_{i,n+1}$ lie on the circle at the intersection of the sphere of radius
$|k_{i,n}|$ defined by Eq.~(\ref{eq:sphere}) and the plane defined
by Eq.~(\ref{eq:plane}). We therefore discuss the splitting in terms
of the vector $\mathbf{k}_{n+1}=(k_{1,n+1},k_{2,n+1},k_{3,n+1})$.
The tip of the vector lies on a circle, hence we parametrise it with
an angle $\theta_{n}$ as:\begin{equation}
\mathbf{k}_{n+1}=k_{n}\left(\mathbf{a}+\sqrt{\frac{2}{3}}\left(\cos\theta_{n}\:\mathbf{u}+\sin\theta_{n}\:\mathbf{v}\right)\right)\:,\label{eq:splitting rule}\end{equation}
where $\mathbf{a}=(1,1,1)/3$, and for instance $\mathbf{u}=(1,-1,0)/\sqrt{2}$
and $\mathbf{v}=\sqrt{3}\:\mathbf{a}\wedge\mathbf{u}$. In fact the choice of  $\theta_{n}$ can be restricted to be picked within the segment $[0,\:2\pi/3]$, since  Eq.(\ref{eq:splitting rule}) exhibits the symmetry  $k_{3,n}(\theta+2/3\pi)=k_{1,n},k_{3,n}(\theta-2/3\pi)=k_{2,n}$ and thus all configurations up to a relabelling can be obtained within this interval.

\subsection{Is the splitting always possible?}

We imagine that we are dealing with a system with many vortices ($2n+1$
after $n$ splittings) and consider the splitting of one point vortex
according to the rules (\ref{geomconditions}) and (\ref{strengconditions})
in this system. The total number of vortices changes, see Fig~\ref{fig:Babies} for an illustration of the process. 

Regarding the
energy we have

\begin{eqnarray}
\delta H_{n} & = & H_{n+1}-H_{n}=-\frac{1}{4\pi}\ln\Lambda_{n}\:,\label{Splitting_energy}\\
 & = & -\frac{1}{4\pi}\left[k_{1}k_{2}\ln R_{3}+k_{1}k_{3}\ln R_{2}+k_{3}k_{2}\ln R_{1}\:,\right]\end{eqnarray}
where for instance the vortex $k_{N}=k_{1}+k_{2}+k_{3}$ was split
in three. In order to be dynamically compatible, we first neglect
the influence of the other vortices (which are considered far enough
to not interfere locally), but still we need to make sure that there
is at least one triangle satisfying the conditions and define the
value of $\delta H_{n}$ associated with the splitting. For that purpose
after the splitting we name $2$ and $3$ the vortices whose strengths
have the same sign with $|k_{2}|<|k_{3}|$, exponentiating Eq.~(\ref{Splitting_energy})
gives\begin{equation}
R_{3}^{k_{1}k_{2}}R_{2}^{k_{1}k_{3}}R_{1}^{k_{2}k_{3}}=\Lambda_{n}\:,\label{eq:Lambda_1}\end{equation}
 then we divide by one noticing that $R_{1}^{\sum k_{i}k_{j}}=1$,
and obtain

\begin{equation}
\left(\frac{R_{3}}{R_{1}}\right)^{k_{1}k_{2}}\left(\frac{R_{2}}{R_{1}}\right)^{k_{1}k_{3}}=\Lambda_{n}\:.\label{eq:Lambda2}\end{equation}
Since the collapse is scale free (self-similar), we choose $R_{1}$
as our length units, note $r_2=R_2/R_1$, $r_3=R_3/R_1$ and arrive at\begin{equation}
r_{2}=\Lambda_{n}^{1/k_{3}k_{1}}r_{3}^{-k_{2}/k_{3}}\:.\label{eq:r2_of_r3}\end{equation}
The condition (\ref{geomconditions}) becomes once rescaled \begin{equation}
k_{1}k_{2}r_{3}^{2}+k_{1}k_{3}r_{2}^{2}+k_{3}k_{2}=0\:,\label{eq:geom_conditions_bis}\end{equation}
combining this last expression with (\ref{strengconditions}) we have
\begin{equation}
k_{2}(r_{3}^{2}-1)+k_{3}(r_{2}^{2}-1)=0\:,\label{eq:geom_condition_quat}\end{equation}
and noting $\alpha=k_{2}/k_{3}(<1)$, we finally obtain\begin{equation}
\alpha r_{3}^{2(1+\alpha)}+\Lambda_{n}^{2/k_{1}k_{3}}=(1+\alpha)r_{3}^{2\alpha}\:.\label{eq:geom_condition_ter}\end{equation}
$\Lambda_{n}$ being positive,  $\alpha X^{1+\alpha}-(1+\alpha)X^{\alpha}$
shows that $r_{3}^{2}\in]0\:;1+1/\alpha[$ and that $0<\Lambda_{n}^{2/k_{1}k_{2}}\le1$.
The equality $\Lambda_{n}=1$ being reached for $r_{3}=1$ which implies
an equilateral triangle. This last configuration has to be excluded
as it is dynamically a fix point, i.e the triangle does not expand
or shrink hence can not be a starting point for a vortex splitting.
So , since $k_{1}k_{3}<0$, this means that $\Lambda_{n}>1$ and that
$\delta H_{n}=-\frac{1}{4\pi}\ln\Lambda_{n}<0$. The splitting of
a vortex in three is thus a dissipative process!

We now enforce that the solution is a triangle, using the rescaled
variables this means that\begin{equation}
1=r_{2}^{2}+r_{3}^{2}-2r_{3}r_{2}\cos\varphi\:,\label{eq:rescaled_triangle}\end{equation}
which implies (using Eq.~(\ref{eq:geom_condition_quat})) \begin{equation}
\cos\varphi=\frac{r_{3}^{2}(1-\alpha)+\alpha}{2r_{3}\sqrt{1+\alpha-\alpha r_{3}^{2}}}\:.\label{eq:Cosinus_triangle}\end{equation}
The right hand side of Eq.~(\ref{eq:Cosinus_triangle}) is always
positive, so $-\pi/2<\varphi<\pi/2$. The minimum
is obtained for $r_{3}^{2}=\alpha(1+\alpha)/(1+\alpha^{2})$ and equal
to $\alpha^{1/2}/(1+\alpha)$, which is always smaller than one. Therefore, there
is always a range of possibilities available for $r_{3}$
and the splitting is always possible. Note that for a given value
of $r_{3}$ the two mirroring shapes of the triangle are possible,
one giving rise to expansion (splitting) the other one to collapse.

\section{Vorticity distributions}

We are now interested in the vorticity distribution we obtain from
such process. For this purpose we compute %
\begin{figure}
\begin{centering}
\includegraphics[width=6cm]{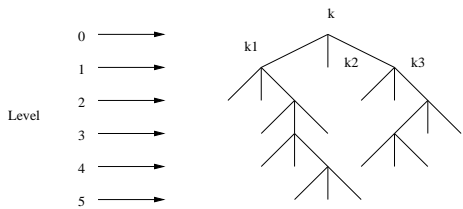} 
\end{centering}
\caption{Simple 5th order vortex lineage. In this picture we see that after $8$ splittings, $4$ vortices are remaining on level $3$.} 

\label{fig:Babies} 
\end{figure}
 different trees originating from one vortex of strength $k=1$ as
depicted in Fig.~\ref{fig:Babies}. The distributions are computed
by successive vortex splitting. After each division, a vortex is chosen
randomly among the global population and is split according to the
reverse collapse rules, the division is the result of the uniformly
random choice of an angle $\theta_{n}$ (see Eq. (\ref{eq:splitting rule})).
Other possible rules made by assigning different probabilities on
vortices will be explored in future work. %
\begin{figure}
\begin{centering}
\includegraphics[width=7cm]{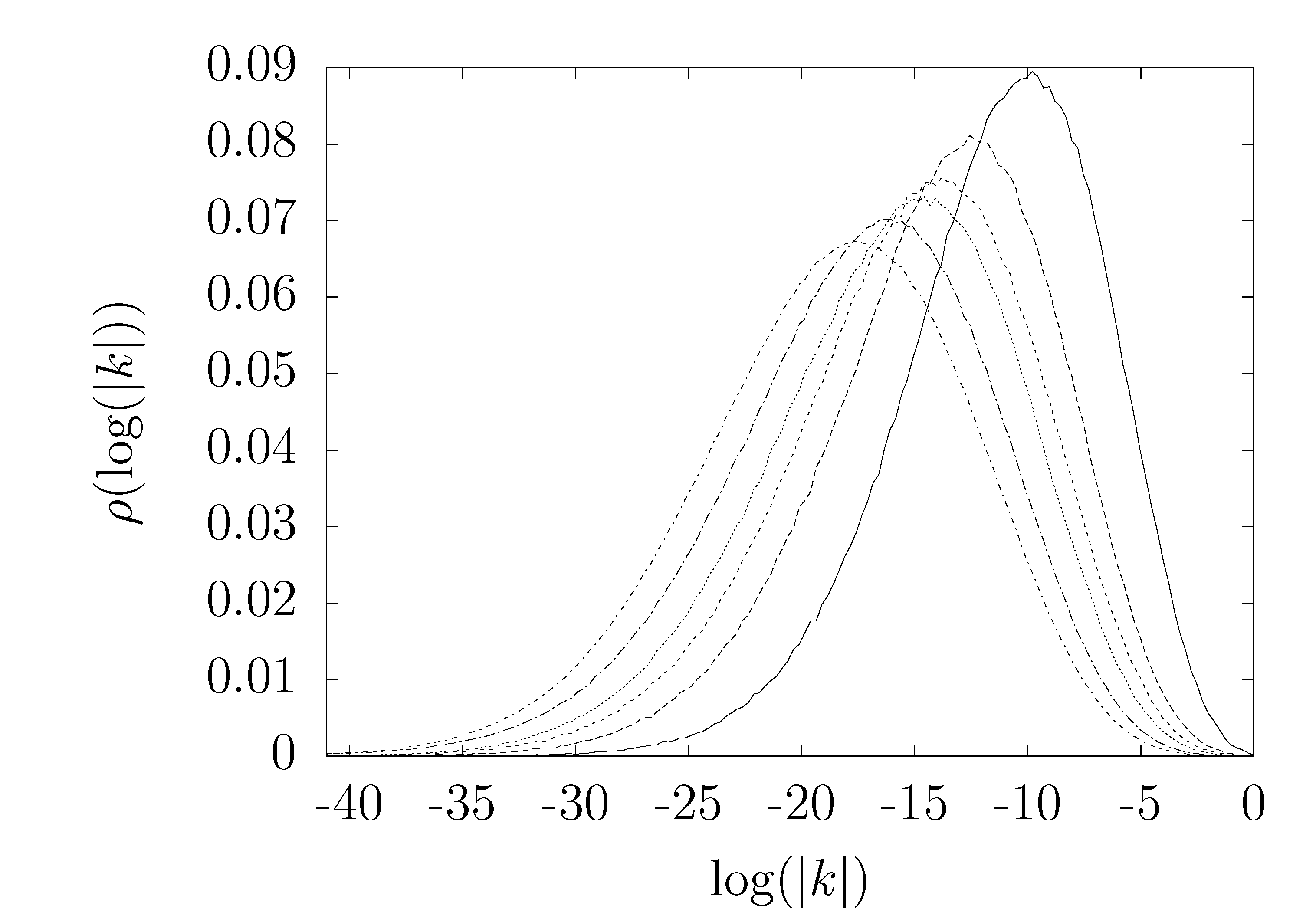} 
\par\end{centering}

\caption{Probability density function (PDF) of the absolutes values of vortex
strengths after respectively (from right to left) $10^{3}$, $5\:10^{3}$,
$10^{4}$, $2\:10^{4}$, $5\:10^{4}$, $10^{5}$ vortex splits. The
PDF have been averaged over $64$ trees, for the large values and
up to $512$ trees for the smallest ones. The shape appears as self-similar.\label{fig:Probability-density-function}}

\end{figure}
Results of the obtained distributions are depicted in Fig.~\ref{fig:Probability-density-function}.
We shall notice that the chosen rules gives rise to a large spectrum
of vorticities (see the logarithmic scale in Fig.\ref{fig:Probability-density-function}).
We notice as well that as the number of division increases the distribution
spreads and the location of its maximum is slowly moving towards smaller
values of the vorticity, moreover, the evolution of the distribution
appears to have some kind of self-similar behaviour. %
\begin{figure}
\begin{centering}
\includegraphics[width=7cm]{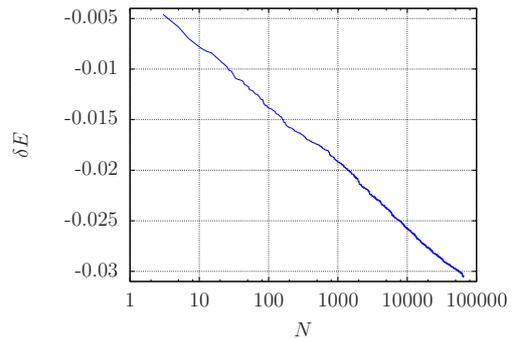} 
\par\end{centering}

\caption{Variation of energy versus number of created vortices. The curve has
been averaged over $100$ trees. Splittings have been performed up
to $N=65537$ vortices starting from one vortex of strength $k_{0}=1$.
One can notice a small logarithmic decay of energy versus the number
of vortices, of the type\noun{ } $\delta E\approx-\lambda\log_{10}(N)$
with $\lambda\approx5.9\:10^{-3}$. \label{fig:energy_vs_N}}

\end{figure}
In the same spirit the variation of energy as a function of generated vortices
can be monitored. Results are shown in Fig.~\ref{fig:energy_vs_N} and show a slow 
logarithmic decay of the total energy.

In order to analyse this in more details we need to characterise the
lineage (see Fig.~\ref{fig:Babies}) after a given number of splittings.
We will note the total number of splittings that occurred $n$. These
successive divisions generate a tree (the phase space of the process)
with $3^{n}$ leaves at the extremities. Each division results in
the choice of an angle $\theta_{i}$. Now let us consider a particular
{}``lineage'' of order $n$, it gives rise to a family of $N=2n+1$
vortices. At each step of the division process (from $n$ to
$n+1$) we choose any already existing vortex and split it with the
rules (\ref{eq:splitting rule}). The trajectory in phase space corresponds
to a connected graph with $2n+1$ leaves starting from the top of
the tree and of total length $3n+1$. The number of possible graphs
on the tree after $n$ splittings is: $(2n+1)!/(2^{n}n!)$. To move
further on, and due to the large amount of possible graphs, we consider
the global occupation $M_{i}$ of the level $i$ (see Fig.~\ref{fig:Babies}).
We note $M_{i}(n)$ the average number of leaves at level $i$ at
time $n$. %
\begin{figure}
\begin{centering}
\includegraphics[width=7cm]{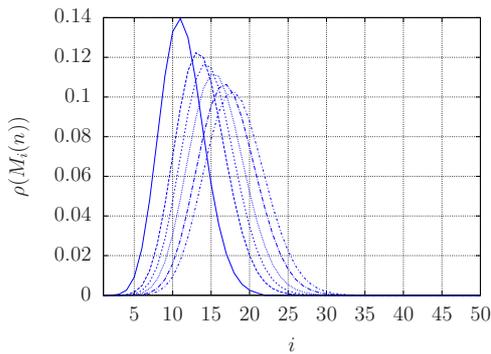} \\
\par\end{centering}

\caption{Probability density function (PDF) of the occupation of each
generation $M$ after respectively (from left to right) $10^{3}$,
$5\:10^{3}$, $10^{4}$, $2\:10^{4}$, $5\:10^{4}$, $10^{5}$ vortex
splits. The PDF have been computed using the mean field equations
(\ref{eq:Occupancy rules}). The shape appears as self-similar and
is reminiscent of what is observed in Fig.~\ref{fig:Probability-density-function}.
\label{fig:Probability-density-function_occupation}}

\end{figure}
Then we have \begin{equation}
M_{i}(n+1)=M_{i}(n)+3\frac{M_{i-1}(n)}{2n+1}-\frac{M_{i}(n)}{2n+1}\:,\label{eq:Occupancy rules}\end{equation}
 with initial conditions $M_{0}(0)=1$, $M_{n>0}(0)=0$. It is easy
to integrate numerically this equation in order to have an idea of
the solution. We find that the form

\begin{equation}
M_{i}(n)=a_{i}\frac{\left(\log(2n+1)\right)^{i-1}}{\sqrt{2n+1}}\label{eq:Occupancy_form}\end{equation}
 is solution, with $a_{i}=3a_{i-1}/(2(i-1))$, i.e. \begin{equation}
a_{i}=\left(\frac{3}{2}\right)^{i-1}\frac{a_{1}}{(i-1)!}\:.\label{eq:prefacteur_occupancy}\end{equation}
This can be checked easily by induction, we assume that $M_{i-1}$
is of this form, and then we can solve Eq.~(\ref{eq:Occupancy rules})
for $M_{i}$ in the continuous time limit in which it becomes \begin{equation}
\frac{dM_{i}}{dn}=3\frac{M_{i-1}}{2n+1}-\frac{M_{i}}{2n+1}\:.\label{eq:Continuous_time_occupancy}\end{equation}
In this way we obtain that at a fixed level $i$, the occupancy $M_{i}$
first increases with time, then decreases, with a maximum at $n^{*}(i)\sim\exp(2i)$.
At fixed time $n$ on the other hand, $M_{i}(n)$ has a maximum at
$i^{*}\sim3\log(2n+1)/2$. %
\begin{figure}
\begin{centering}
\includegraphics[width=7cm]{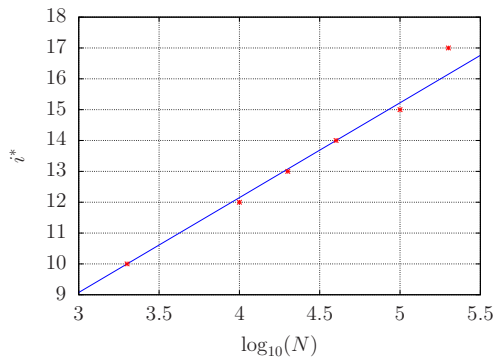} 
\par\end{centering}

\caption{Maximum $i^{*}$ versus number of vortices $N$. $i^{*}$ is obtained
from the distributions displayed in Fig.~\ref{fig:Probability-density-function_occupation}.
A linear scaling is observed as expected, with a measured slope of
$\sim3.1$, {}``close to the expected value $3/2\times\ln10\approx3.45$''
\label{fig:i_star_vs_N}.}

\end{figure}
A numerical integration of the global populations given by Eq.~(\ref{eq:Occupancy rules}) is displayed in Fig.~\ref{fig:Probability-density-function_occupation}. One can notice similarities with the distributions of
vorticity although the distributions are more peaked.  In order to test as well our analysis, the location of the maximum of the distribution is displayed in Fig.~\ref{fig:i_star_vs_N} and a good agreement with the logarithmic law is found.

Let us now compute the distribution of vorticity $\rho(k,n)$ assuming we know
the occupancy $M_{i}(n)$. Hence let us consider a vortex living in
the generation $i$. It has been the result of $i$ splitting. Since
the splitting rules (\ref{eq:splitting rule}) have no preferred order
(as mentioned earlier they permute if we add $2\pi/3$ to the random
angle), we assume that the obtained vortex is always the third vortex
hence its absolute vorticity $k$ will end up being \begin{equation}
k=\frac{1}{3^{i}}\prod_{j=1}^{i}(1-2\sin\theta_{j})\:,\label{eq:possible_k}\end{equation}
and consequently its logarithm is \begin{equation}
\log|k|=\sum_{j=1}^{i}\log|1-2\sin\theta_{j}|-i\log3\:,\label{eq:possible_logk}\end{equation}
with $\theta_{j}$ being uniformly distributed random variables in
$[0\:2\pi[$. We can then gather the probability distribution of vortex
strengths at generation $i$, which we note $\rho_{i}(k)$. And thus
the vorticity distribution after $N$ division writes \begin{equation}
\rho(k,n)=\sum_{i}\rho_{i}(k)\rho(M_{i},n)\:,\label{eq:distribution_of_k}\end{equation}
 with $\rho(M_{i},n)=M_{i}/(2n+1)$.%
\begin{figure}
\begin{centering}
\includegraphics[width=7cm]{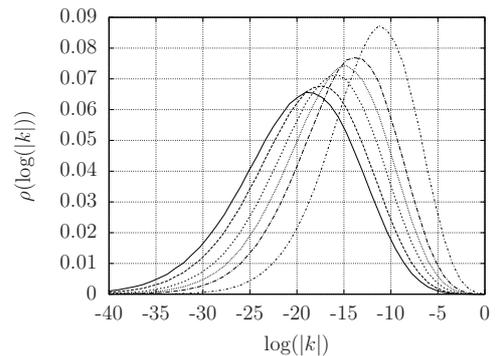} 
\par\end{centering}

\caption{Probability density function (PDF) of the absolutes values of vortex
strengths after respectively (from right to left) $10^{3}$, $5\:10^{3}$,
$10^{4}$, $2\:10^{4}$, $5\:10^{4}$, $10^{5}$ vortex splits. The
PDF have been computed using Eq.(\ref{eq:distribution_of_k}), where
we took into account only up to generation $i=40$. The results appear
to be coherent with what is displayed in Fig.~\ref{fig:Probability-density-function}.\label{fig:Probability-density-function-1}}

\end{figure}

In order to check these results, we compute $\rho(k,n)$ using Eq.(\ref{eq:distribution_of_k})
and compare the results to those displayed in Fig.~\ref{fig:Probability-density-function}.
In fact, given the shape of the occupancy $M_{i}(n)$, we can compute
$\rho(k,n)$ using only a {}``few'' distributions $\rho_{i}(k)$.
For instance in Fig.~\ref{fig:Probability-density-function-1} we
computed the $\rho_{i}(k)$ taking into account in the tree the vortices only up to generation $i=40$. We notice as well a very similar behaviour as the one displayed in Fig.~\ref{fig:Probability-density-function}.

\section{Conclusion}

This paper is a first attempt at analysing the distribution of vorticities
originating from one point vortex using a dynamically compatible process,
namely a reverse collapse route with the conservation of total vorticity.
The splitting process has been analysed in great details and shown
to be dissipative. Afterwards a simple process consisting of randomly
successive splittings is proposed and the resulting distributions
have been analysed. Analytical computation of the proposed process
have been made, resulting for instance in the computation of the vorticity
distribution after $n$ consecutive splitting of vortices and show
very good agreement with the numerical simulation of the process.
This paper is  a first step for further work. One could for
instance modify the splitting rules in order to obtain a conservative
process, but also could try to pick vortices and change how the splitting
is done, with a non uniform probabilities, such as a Gibbsian one
and analyse the resulting distributions. In other words, we could perform statistical
physics of point vortices allowing a varying number of vortices according
to the collapse rules, and see if this possibility changes the equilibrium
features. Last but not least, it will be important to compare
the obtained distribution with real data involving physical stochastic processes. In particular, it is tempting to compare
the vortex distribution obtained with this proposed mechanism
or its variants with experimental results obtained on two-dimensional
physical flows. Remarkably, one has to notice that, starting
with a positive vortex, the reverse collapse course induces naturally
the creation of negative vortices, thus the engendered distributions
will consist of both positive and negative vortices. Work is currently
under way to analyse these different possibilities.

\acknowledgement{
X. Leoncini and S. Villain-Guillot thank Soci\'et\'e Math\'ematique de Paris Foundation for support during their attendance at the trimester ``Singularities in mechanics'' held at I.H.P during the first trimester of 2008,  where
first discussion about this work took place. We would like to thank A. Verga for useful discussions and comments.}

\bibliographystyle{epj}

\begin{thebibliography}{99}
\bibitem{frisch} U. Frisch "Turbulence: the legacy of A.N. Kolmogorov", Cambridge Univ. Press (1995).

\bibitem{Benzi86}
R.~Benzi, G.~Paladin, S.~Patarnello, P.~Santangelo, A.~Vulpiani, J. Phys. A
  \textbf{19}, 3771 (1986)

\bibitem{Benzi88}
R.~Benzi, S.~Patarnello, P.~Santangelo, J. Phys. A \textbf{21}, 1221 (1988)

\bibitem{McWilliams84}
J.C. McWilliams, J. Fluid Mech. \textbf{146}, 21 (1984)

\bibitem{Weiss92}
J.B. Weiss, J.C. McWilliams, Phys. Fluids A \textbf{5}, 608 (1992)

\bibitem{Carnevale91}
C.F. Carnevale, J.C. McWilliams, Y.~Pomeau, J.B. Weiss, W.R. Young, Phys. Rev.
  Lett. \textbf{66}, 2735 (1991)

\bibitem{Zabusky82}
N.J. Zabusky, J.C. McWilliams, Phys. Fluids \textbf{25}, 2175 (1982)

\bibitem{Vobseek97}
P.W.C. Vobseek, J.H.G.M. {van Geffen}, V.V. Meleshko, G.J.F. {van Heijst},
  Phys. Fluids \textbf{9}, 3315 (1997)

\bibitem{Fuentes96}
O.U. {Velasco Fuentes}, G.J.F. {van Heijst}, N.P.M. {van Lipzig}, J. Fluid
  Mech. \textbf{307}, 11 (1996)

\bibitem{Benzi92}
R.~Benzi, M.~Colella, M.~Briscolini, P.~Santangelo, Phys. Fluids A \textbf{4},
  1036 (1992)

\bibitem{Weiss98}
J.B. Weiss, A.~Provenzale, J.C. McWilliams, Phys. Fluids \textbf{10}, 1929
  (1998)

\bibitem{Dritschel96}
D.G. Dritschel, N.J. Zabusky, Phys. Fluids \textbf{8}, 1252 (1996)

\bibitem{Aref99}
H.~Aref, \emph{Turbulent Statistical dynamics of a system of point vortice}
  (Birkh{\"a}user Verlag, 1999), p. 151, Trends in Mathematics, ISBN
  978-3-7643-6150-1

\bibitem{Onsager49}
L.~Onsager, Nuovo Cimento, Suppl. \textbf{6}, 279 (1949)

\bibitem{Joyce73}
G.~Joyce, D.~Montgomery, J. Plasma Phys. \textbf{10}, 107 (1973)

\bibitem{frohlich82}
J.~Fr{\"o}hlich, D.~Ruelle, Commun. Math. Phys. \textbf{87}, 1 (1982)

\bibitem{Caglioti92}
E.~Caglioti, P.L. Lions, C.~Marchioro, M.~Pulvirenti, Commun. Math. Phys.
  \textbf{143}, 501 (1992)

\bibitem{Robert92}
R.~Robert, J.~Sommeria, Phys. Rev. Lett. \textbf{69}(19), 2776 (1992)

\bibitem{Spineanu05}
F.~Spineanu, M.~Vlad, Phys. Rev. Lett. \textbf{95}(23), 235003 (2005)

\bibitem{Chavanis07}
P.H. Chavanis, M.~Lemou, Eur. Phys. J. B \textbf{59}, 217 (2007)

\bibitem{Synge49}
J.L. Synge, Can. J. Math. \textbf{1}, 257 (1949)

\bibitem{Novikov79}
E.A. Novikov, Y.B. Sedov, Sov. Phys. JETP \textbf{22}, 297 (1979)

\bibitem{LKZ2000}
X.~Leoncini, L.~Kuznetsov, G.M. Zaslavsky, Phys. Fluids \textbf{12}, 1911
  (2000)

\bibitem{Novikov75}
E.A. Novikov, Sov. Phys. JETP \textbf{41}, 937 (1975)

\bibitem{Aref79}
H.~Aref, Phys. Fluids \textbf{22}, 393 (1979)

\bibitem{Aref83}
H.~Aref, Ann. Rev. Fluid Mech. \textbf{15}, 345 (1983)

\bibitem{Tavantzis88}
J.~Tavantzis, L.~Ting, Phys. Fluids \textbf{31}, 1392 (1988)

\end{thebibliography}

\end{document}